\documentclass[12pt,draftcls,onecolumn]{IEEEtran} 

\usepackage{amsfonts}
\usepackage{amssymb}
\usepackage{amsbsy} 
\usepackage{amsmath}
\usepackage{subfigure}
\usepackage[dvips]{graphicx}
\usepackage[T1]{fontenc}
\usepackage[latin1]{inputenc}
\usepackage{color}
\usepackage{array}
\usepackage{graphicx}
\usepackage{threeparttable}

\usepackage[font=small,skip=0pt]{caption}

\usepackage{hyperref}
\hypersetup{
    bookmarks=true,         
    unicode=false,          
    pdftoolbar=true,        
    pdfmenubar=true,        
    pdffitwindow=false,     
    pdfstartview={FitH},    
    pdftitle={IEEE GRSM},    
    pdfauthor={Camps-Valls},     
    pdfsubject={GPR survey},   
    pdfcreator={Camps-Valls},   
    pdfproducer={Camps-Valls}, 
    pdfkeywords={keyword1} {key2} {key3}, 
    pdfnewwindow=true,      
    colorlinks=true,       
    linkcolor=blue,          
    citecolor=blue,        
    filecolor=blue,      
    urlcolor=blue           
}

\newcommand{\IG}{\includegraphics}
\usepackage{multicol}
\usepackage{float}
\usepackage{lscape}
\usepackage{cite}

\usepackage[usenames,dvipsnames]{xcolor}
\definecolor{mycolor}{rgb}{0.1, 0.5, 0.2}

\renewcommand{\baselinestretch}{0.9794} 

\begin{document}

\title{Active Learning Methods for Efficient Hybrid Biophysical Variable Retrieval}

\author{Jochem Verrelst, Sara Dethier, Juan Pablo Rivera, Jordi Mu\~noz-Mar\'i, \\
Gustau Camps-Valls, and Jos\'e Moreno
\thanks{\bf Preprint. Paper published in IEEE Geoscience and Remote Sensing Letters, vol. 13, no. 7, pp. 1012-1016, July 2016, doi: 10.1109/LGRS.2016.2560799.}
\thanks{This work was partially supported by the Spanish Ministry of Economy and Competitiveness under project TIN2015-64210-R and the European Space Agency (ESA) projects 'FLEX/S3 Tandem Mission Performance Analysis and Requirements Consolidation Study (PARCS)' (RFQ 3-13397/11/NL/CBi) and 'FLEX-Bridge Study' (RFP IPL-PEO/FF/lf/14.687).}
\thanks{JV, JPR, JMM, GCV and JM are with the Image Processing Laboratory (IPL). Universitat de Val{\`{e}}ncia. C/ Catedr\'atico Escardino, Paterna (Val{\`{e}}ncia) Spain. 
Web: http://isp.uv.es. E-mail: \{jochem.verrelst,juan.rivera,jordi.munoz,gustau.camps,jose.moreno\}@uv.es.\newline
\indent SD is with Department of Physics, Imperial College London, UK. E-mail: sara.dethier11@imperial.ac.uk }

}

\markboth{IEEE Geoscience and Remote Sensing Letters, Vol. XX, No. Y, Month Z 2016}
{Verrelst et al.: Active learning for retrieval}

\maketitle

\begin{abstract}
Kernel-based machine learning regression algorithms (MLRAs) are potentially powerful methods for being implemented into operational biophysical variable retrieval schemes. However, they face difficulties in coping with large training datasets. With the increasing amount of optical remote sensing data made available for analysis and the possibility of using a large amount of simulated data from radiative transfer models (RTMs) to train kernel MLRAs, efficient data reduction techniques will need to be implemented. Active learning (AL) methods enable to select the most informative samples in a dataset. This letter introduces six AL methods for achieving optimized biophysical variable estimation with a manageable training dataset, and their implementation into a Matlab-based MLRA toolbox for semi-automatic use. The AL methods were analyzed on their efficiency of improving the estimation accuracy of leaf area index and chlorophyll content based on PROSAIL simulations. Each of the implemented methods outperformed random sampling, improving retrieval accuracy with lower sampling rates. Practically, AL methods open opportunities to feed advanced MLRAs with RTM-generated training data for development of operational retrieval models.

\end{abstract}



\section{Introduction}\label{sec:intro}
Spatiotemporal quantitative retrieval methods for the Earth's surface are a requirement in a variety of Earth system applications. Optical Earth observing satellites, endowed with high temporal resolutions, enable the retrieval and hence monitoring of bio-geophysical variables~\cite{schaepman09}. With the forthcoming super-spectral Copernicus Sentinel-2 and Sentinel-3 missions, as well as the planned hyperspectral EnMAP, HyspIRI, PRISMA and ESA's candidate FLEX missions, an unprecedented data stream for land monitoring will soon become available to a diverse user community. This vast data stream will require enhanced processing techniques that are accurate, robust and fast.

One of the major challenges in this respect is to cope with the large amount of data that has to be processed. Over the last few decades a wide diversity of biophysical variable retrieval methods have been developed, but only a few of them made it into operational use, and many of them are still in their infancy~\cite{Verrelst2015b}. Generally, there are four main approaches to the problem of estimating biophysical variables from reflectance spectra, i.e. (1) parametric regression, (2) non-parametric regression, (3) physically-based, i.e. inverting radiative transfer models (RTMs), and (4) hybrid methods. Hybrid methods combine the generic properties of physically-based models combined with the flexibility and computational efficiency of non-parametric, non-linear regression models ~\cite{Verrelst2015b}. Hybrid retrieval schemes proved to be successful in operational processing chains. Current hybrid schemes for the generation of land products such as leaf area index (LAI) rely on neural networks (NNs), typically trained using a very large amount of RTM-simulated data ~\cite{Baret07,Baret2013}. For LAI retrievals for instance, the most widely used RTM is PROSAIL~\cite{Jacquemoud09}. 

Apart from NNs, for the last decade various alternative non-parametric methods in the field of machine learning regression algorithms (MLRAs) have been introduced, many of them with interesting properties. Particularly, kernel-based MLRAs, such as kernel ridge regression (KRR) ~\cite{Suykens1999}, have proven to be simple for training and yield competitive accuracy~\cite{Verrelst12a}. Gaussian process regression (GPR) can even provide associated uncertainty intervals for the estimations~\cite{Rasmussen06,Verrelst2012}. They both could become powerful alternatives to replace NNs in hybrid retrieval schemes~\cite{Verrelst2015}. However, these advanced MLRAs go at a computational cost, typically scaling poorly with the number of samples. This is currently the main shortcoming for their adoption in hybrid schemes for retrieval.

The problem when using RTM-simulated data is that these data can easily scale up to several millions of samples corresponding to different variable selections, especially when look-up tables (LUTs) are used and all input variables are systematically combined. Simulated data created in this way might be redundant, as slight variable variations might produce very similar spectra. 
Therefore, in order to have these kernel-based MLRAs trained by large datasets, we need intelligent sampling approaches that select the most informative samples from a large pool dataset. In the spectral domain we can use feature extraction techniques~\cite{Arenas2013} to reduce these large datasets. In the sampling domain, we propose to use AL techniques~\cite{MacKay1992,Tuia11} to obtain an optimized training set while keeping its size small enough. This letter analyzes the utility of those AL techniques for the benefit of developing efficient hybrid retrieval schemes.

This brings us to the following main objective: to introduce AL techniques that deal efficiently with RTM-simulated data to optimize kernel-based MLRA hybrid retrieval strategies. The sub-objectives are: (1) to implement AL methods into a software framework that facilitates the use of these methods into hybrid retrieval strategies, and (2) to evaluate the efficiency of AL methods in optimizing hybrid PROSAIL-MLRA models with a manageable training dataset. This analysis is illustrated for the retrieval of LAI and leaf chlorophyll content (LCC) from simulated Sentinel-3 OLCI data.

\section{Active Learning theory}\label{sec:AL}
Let us assume that we have a large (simulated) training dataset composed of many samples, so large that it is unfeasible to use all the samples to obtain a regression model. In order to reduce its size, the typical naive approach is to select some samples from it randomly. This strategy does not necessarily optimize the selection since all samples, including redundant and noisy samples, are treated equally. AL ~\cite{MacKay1992} can be used to solve this shortcoming using smart sample selection criteria to improve data compactness, diversity and richness, and hence the model's estimation accuracy. AL techniques aim to select the most informative samples in a training dataset to reach high accuracy using fewer labelled samples.
While widely used in data classification ~\cite{Tuia11,Jordi2012,Crawford2013,Pasolli2014}, there are few experimental studies carried out for regression problems, and these are focused on the estimation of terrestrial ~\cite{Pasolli2012} and ocean biophysical~\cite{Bazi07,Douak2013} variables from remote sensing data. None of these aims to be generic at larger scales via the training of RTM simulations. Accordingly, the application of AL techniques to smartly sample large simulated training datasets in hybrid schemes creates the opportunity to improve the prediction accuracy of the resulting regression model. 
AL methods start with a small training set of labelled data - in this case reflectance-variable pairs - and use query strategies, or selection criteria, to select samples from a larger unlabelled data pool, i.e. one in which samples have no assigned variable value. The selected samples are labelled and are added to the training set until it becomes optimal~\cite{Douak2013}. In classification the samples are typically labelled by a human expert. The use of simulated training data removes the need for a human expert as the labels are taken from the LUT, thus creating an automated optimized sampling solution.  
The optimal final training set should be small enough to substantially increase the computational efficiency of the final prediction model, and large enough to accurately represent the original training set and give highly accurate predictions. The selection criteria select the most significant training samples from the pool, i.e. those which if labelled would improve the regression model the most.

The selected criterion algorithms can rank the samples according to the \textit{uncertainty} of a sample or its \textit{diversity}~\cite{Crawford2013}. %
These criteria are sometimes used together within classification problems~\cite{Gu2015}, and are here applied separately to regression. The separate use of these criteria in this study already provides benefits, although it is possible to combine them as in classification problems.
Selecting samples by uncertainty picks the most uncertain samples, i.e. those with the least confidence. Uncertainty criteria include variance-based pool of regressors (PAL)~\cite{Douak2013}, entropy query-by-bagging (EQB)~\cite{Tuia11}, residual regression (RSAL)~\cite{Douak2011}. Selecting samples by diversity ensures that added samples are dissimilar from those already accounted for. Diversity criteria include Euclidean distance-based diversity (EBD)~\cite{Douak2013}, angle-based diversity (ABD)~\cite{Demir2011} and cluster-based diversity (CBD)~\cite{Patra2012}. The algorithms are briefly described below. 

\subsection{Uncertainty criteria methods}\label{uncertainty}
 
\paragraph{Variance-based Pool of Regressors (PAL)}

This strategy~\cite{Douak2013} first generates $k$ subsets by randomly choosing samples from the original training set. Each subset is then used to train a regressor and to obtain a prediction for each sample in the candidate set. This ends up with $k$ different predictions for each candidate sample. Then, the variance of each prediction is estimated as:
\begin{equation}
	\sigma_y^2 = \frac{1}{k}\sum_{i=1}^{k}(y_i - \bar{y})^2,
\end{equation}
where $\bar{y}=\frac{1}{k}\sum_{i=1}^k{y_i}$. The variance gives an indication of the spread of the estimations. The samples with the highest variance, i.e. greater disagreements between the different regressors, are added to the training set and removed from the candidate set.

\paragraph{Entropy Query-by-Bagging (EQB)}

In this approach~\cite{Tuia11}, the predictions of $k$ different regressors are ranked according to their entropy:
\begin{equation}
	H(x) = - \sum_{i=1}^k p(x_i)\log{p(x_i)},
\end{equation}
where $p(x_i)$ is the probability of the sample $x$ being predicted by the regressor $i$. Samples for which various regressors give similar values have lower uncertainties, and this is indicated by smaller or more negative entropy values. $H(x)$ is computed for each sample and then the samples with the greatest entropy are added to the training set.

\paragraph{Residual Regression AL (RSAL)}

Adapted from~\cite{Douak2011}, this method quantifies the systematic errors generated by a regression algorithm. It does so by training a second model (residual model), which estimates the prediction errors, $e(x) = y-\hat{y}$, where $y$ is the actual observed value, and $\hat{y}=\hat{f}(x)$ is the model prediction given the input $x$. The algorithm selects the samples that exhibit a high prediction error and adds these to the training set.

\subsection{Diversity criteria methods}\label{diversity}

\paragraph{Euclidean Distance-Based Diversity (EBD)}

This method~\cite{Douak2013} selects the samples in the candidate set that are distant from the current training set using their squared Euclidean distance:
\begin{equation}
	d_E = \|x_u - x_l\|_2^2,
\end{equation}
where $x_u$ is a sample from the candidate set, and $x_l$ is a sample from the training set. All distances between samples are computed and then the farthest are selected.

\paragraph{Angle Based Diversity (ABD)}

This strategy~\cite{Demir2011} measures the diversity between samples using the cosine angle distance, defined as:
\begin{equation}
	\angle(x_u,x_l) = \cos^{-1}\left( \frac{\langle x_u,x_l \rangle}{\|x_u\|\cdot\|x_l\|} \right)
\end{equation}
where $\langle x_u,x_l \rangle$ is the inner product between $x_u$ and $x_l$.  
The learning samples showing largest cosine angles with the training data
are added to the training set.

\paragraph{Cluster Based Diversity (CBD)}

This method~\cite{Patra2012} first groups the data using a clustering algorithm, for instance $k$-means. The number of clusters $k$ is set to the number of samples to add in each iteration of the AL algorithm. For each cluster, the nearest sample to the cluster centroid is selected.

\section{ARTMO toolbox}\label{sec:ARTMO}
This study was conducted within an in-house developed graphical user interface (GUI) software package named ARTMO (Automated Radiative Transfer Models Operator)~\cite{Verrelst12c}. 
ARTMO embodies a suite of leaf and canopy radiative transfer models (RTMs) including PROSAIL (i.e. the leaf model PROSPECT coupled with the canopy model SAIL~\cite{Jacquemoud09}) and several retrieval toolboxes, i.e. a spectral indices toolbox, a LUT-based inversion toolbox, and a machine learning regression algorithm (MLRA) toolbox. 

The MLRA retrieval toolbox~\cite{Rivera13b} offers a suite of regression algorithms that enable the estimation of biophysical variables based on either experimental or simulated data with little user interaction. The toolbox is built around the SimpleR package~\cite{simpler} with over 15 MLRAs, and additionally includes feature reduction techniques such as principal component analysis (PCA) and non-linear alternatives (e.g. kernel-PCA) along with cross-validation sub-sampling techniques. The option of applying AL methods was implemented for the first time in the latest MLRA toolbox version (v1.16). A GUI has been developed that allows selecting one or multiple AL methods, as well as the selection of a starting dataset, the number of samples to add per iteration and the stopping criterion. A goodness-of-fit validation overview table allows the user to select the best-performing hybrid method, which then can be used to process remote sensing images efficiently. The ARTMO package runs in MATLAB and can be freely downloaded from~\url{http://ipl.uv.es/artmo/}.

\section{Experimental setup}\label{sec:expsetup}
\subsection{RTM simulations}\label{RTMsim}

PROSAIL-simulated spectral and input variable data was used to test and compare the AL techniques on different kernel regression models. Sensor settings corresponding to S3-OLCI were chosen over 18 bands from 443 to 940 nm. Different selections of variables such as solar angles and leaf characteristics were selected in order to create a generic dataset by combining the ranges of all variables, as shown in  Table~\ref{LUTconf}. 
The resulting dataset contained 218,750,000 PROSAIL-simulated reflectance-parameter pairs and was stored in a look-up table in ARTMO. This dataset, which is is overly big and largely redundant for regression, was randomly sampled into subsets of different sizes ranging between 1000 and 100,000 samples. The various subsets were used to investigate when the computational processing of the kernel regressions failed, using a random training/validation splitting of 50\%. 

\begin{table}[!ht] 
\begin{center}
    \caption{Range and distribution of input variables used to establish the synthetic canopy reflectance database for use in the LUT.}
    \resizebox{1\textwidth}{!}{ 
     \begin{tabular}{llccc@{}}
\hline
       \multicolumn{2}{c}{\bf Model variables} & {\bf Units} & {\bf Range} & {\bf Distribution}\\
\hline			
\multicolumn{5}{l}{$Leaf$ $variables$: PROSPECT-4} \\
$N$ &  Leaf structure index & unitless & 1.3-2.5 & Uniform \\
LCC & Leaf chlorophyll content & [${\mu}$g/cm$^{2}$]  & 5-75 & Gaussian (${\bar x}$: 35, SD: 30) \\
$C_m$ &  Leaf dry matter content & [g/cm$^{2}$] & 0.001-0.03 & Uniform\\
$C_w$ &  Leaf water content & [cm] & 0.002-0.05 & Uniform\\
\\
\multicolumn{5}{l}{$Canopy$ $variables$: 4SAIL}\\
LAI & Leaf area index & [m$^{2}$/m$^{2}$] & 0.1-7 & Gaussian (${\bar x}$: 3, SD: 2)\\
$\alpha_{soil}$ & Soil scaling factor & unitless & 0-1 & Uniform \\
ALA & Average leaf angle & [$\circ$] & 40-70 & Uniform\\
HotS & Hot spot parameter & [m/m] & 0.05-0.5 & Uniform\\
skyl & Diffuse incoming solar radiation & [fraction] & 0.05 & - \\     
$\theta_s$ & Sun zenith angle & [$\circ$] & 22.3 & - \\
$\theta_v$ & View zenith angle & [$\circ$] & 20.19 & - \\
$\phi$ & Sun-sensor azimuth angle & [$\circ$] & 0 & - \\ 
\hline
	   \end{tabular}}
	   \label{LUTconf}
 \begin{tablenotes}
 \small
\item Similar variable ranges/values/distributions were used according to field configurations and related studies ~\cite{Rivera13a}. ${\bar x}$: mean, SD: standard deviation. 
\end{tablenotes}
\end{center}
\end{table}

\subsection{MLRA configuration}\label{MLRA}

Two kernel-based MLRAs were selected: KRR and GPR. These methods have been amply described before (e.g.~\cite{Verrelst2015b}) and are here only briefly summarized.  

KRR minimizes the squared residuals in a higher dimensional feature space, and can be considered as the kernel version of the regularized ordinary least-squares linear regression~\cite{Scholkopf02}. The linear regression model is defined in a Hilbert space, ${\mathcal H}$, of very high dimensionality, where samples have been mapped to through a mapping $\boldsymbol{\phi}({\bf x}_i)$. We have used the radial basis function (RBF) kernel function.

GPR has recently been introduced as a powerful regression tool~\cite{Rasmussen06}. The model provides a probabilistic approach for learning generic regression problems with kernels. The GPR model establishes a relation between the input and the output variable in the same way as KRR. However, two additional advantages of GPR must be noted. First, not only a predictive mean but also a predictive variance can be obtained, which can serve as an uncertainty interval. The second advantage is that one can use very sophisticated kernel functions  because hyperparameters can be learned efficiently by maximizing the marginal likelihood in the training set ~\cite{Rasmussen06,Verrelst2012}. In the current study a standard RGB kernel function was used.

\subsection{AL sampling design}\label{ALsample}

Although KRR and GPR proved to deliver high accuracies and were fast in processing, they incurred a computational cost, thus limiting the number of input training samples. 
Initial tests revealed that the largest set that could be processed under a reasonable amount of time for both regression algorithms was the set containing 5000 samples of simulated RTM data, thus this was used to analyze the AL methods.

The 5000 samples were separated into an initial learning set of 2500 samples and a validation set of 2500 samples. The GPR and KRR models were first trained and validated on those datasets, further referred to as full training models, and were used as reference models. Subsequently, an initial random training set of 50 samples of spectral data was selected from the learning set, followed by MLRA training and validation. Samples from the learning set were then added to the training set using the AL methods in batches of 50 samples per iteration until no more samples were left in the learning set. The labels at each iteration were taken from the LUT containing the simulated reflectance-parameter pairs.
This was repeated 10 times with a different initial training set each time to reach statistically reliable results. 
The number of pools used in the PAL and EQB methods ($k$) were determined by comparing their predictive capability and processing time using 5 and 10 pools in the initial testing phase. Since they both gave similar results, the number of pools was fixed to 5 in order to be faster.
Finally, the averaged coefficients of determination ($R^2$) values of the 10 runs over the first 19 iterations, i.e. to reach 1000 training samples, were plotted. This procedure was completed using KRR and GPR to retrieve LAI and LCC values, though because of leading to similar results only LCC by KRR and LAI by GPR results are addressed here.

\section{Results}\label{sec:results}

\begin{figure}[H]
 \centering
	\footnotesize
	\begin{tabular}{cc}
KRR to LCC  & GPR to LAI \\ 
\IG[width=8.0cm]{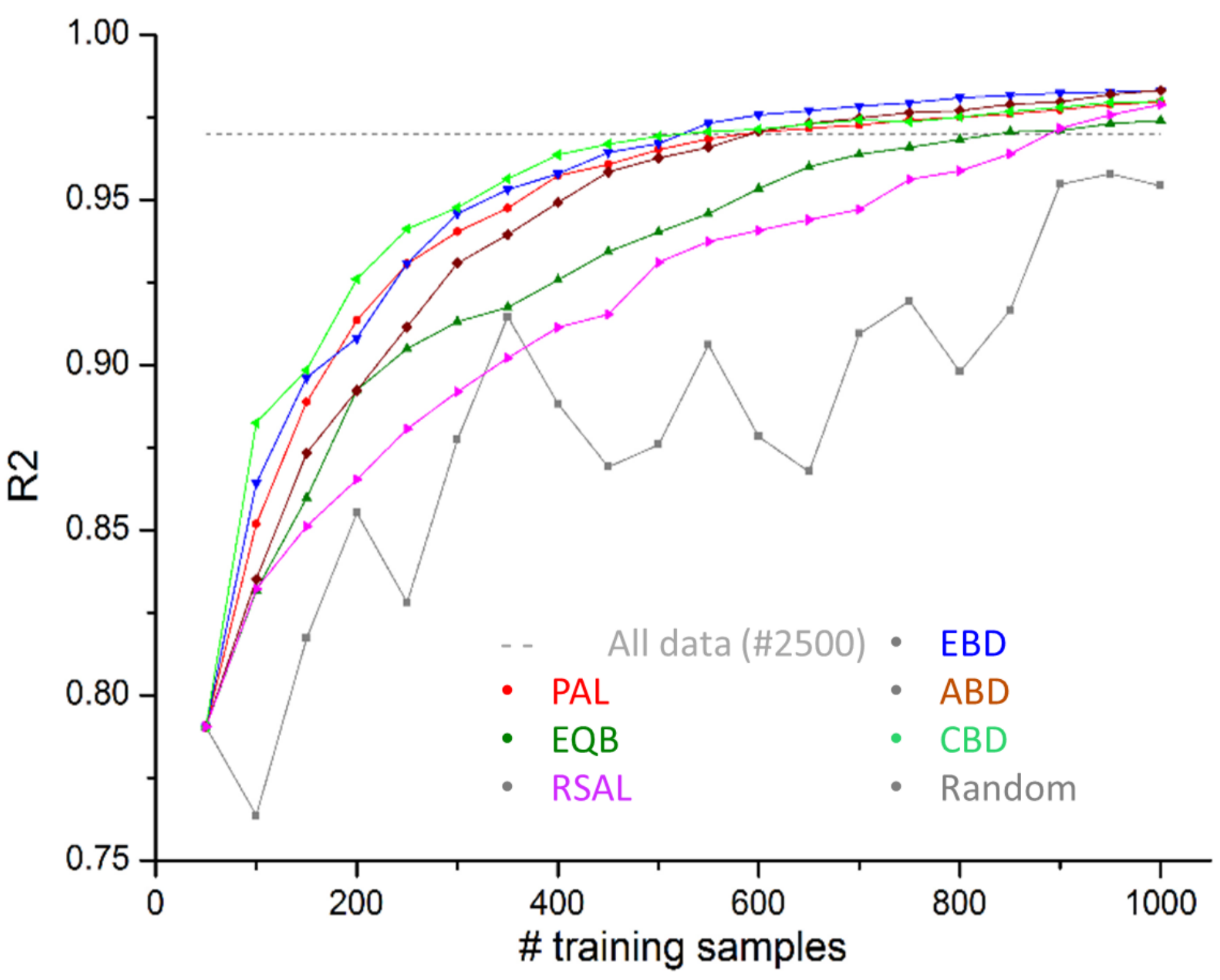}  & \IG[width=8.0cm]{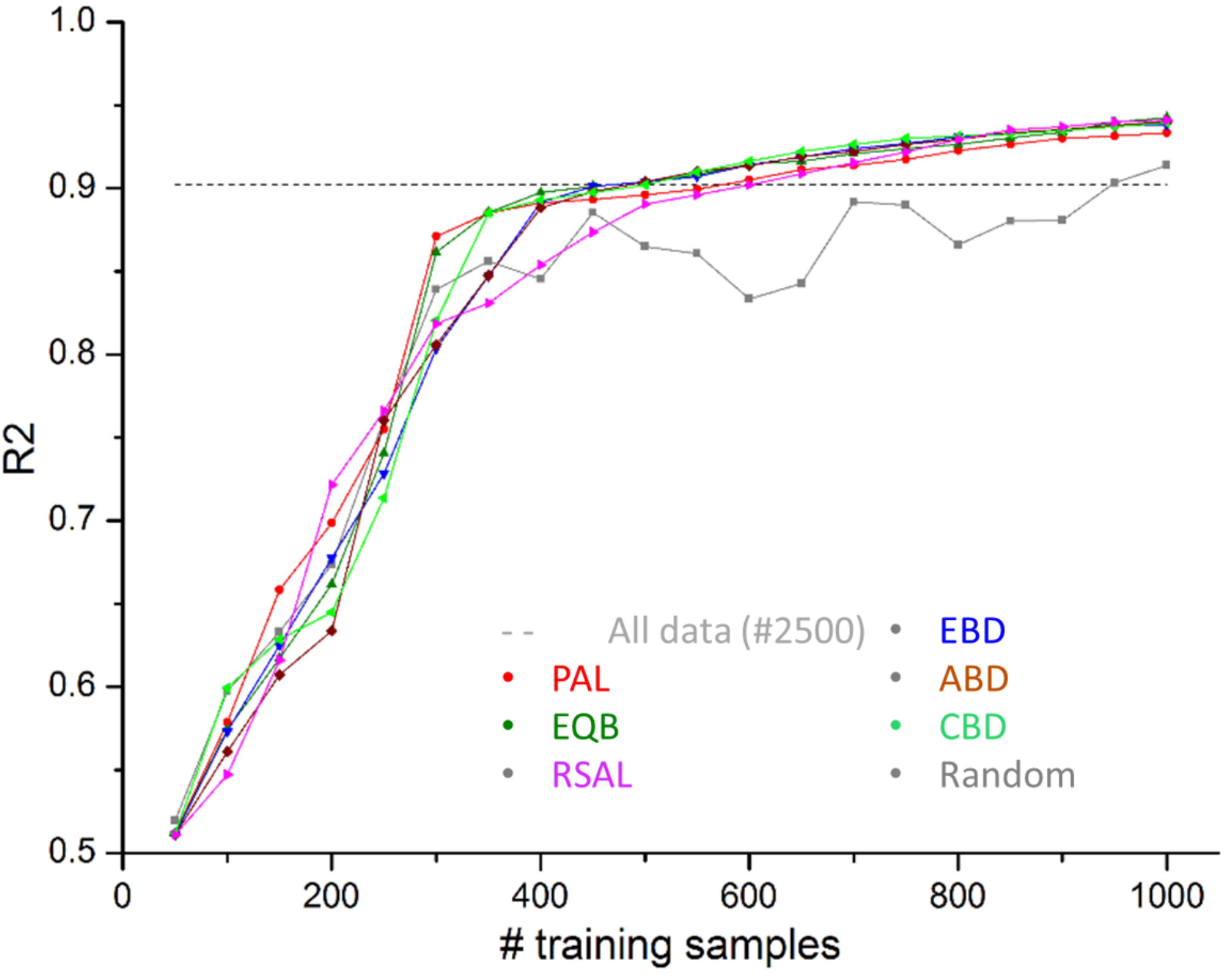} \\

\end{tabular}\vspace{12pt}
 \caption{Mean $R^2$ using AL methods with KRR for LCC estimation [left] and GPR for LAI estimation [right].}
 \label{AL_results}
\end{figure}

Figure \ref{AL_results} shows the average $R^2$ performances of the AL methods over 10 runs against validation data, each time until a training set of 1000 is reached.  
The dashed line represents the $R^2$ accuracy of the full reference training dataset containing 2500 samples. The grey line represents the performance of the random sampling sequence, which is added to appreciate the benefit of AL. The colored lines represent the six AL methods considered. For both LCC and LAI, the random sampling technique reduces the uncertainty in the prediction model as more samples are added, but it is unstable as it leads to irregular convergence curves and is outperformed by the AL methods in terms of convergence rate. All AL methods surpass the reference training dataset before reaching 1000 training samples.
 
Regarding LCC retrieval with KRR, the diversity AL methods EBD, ABD and CBD, along with the uncertainty method PAL led to similarly good results. These methods reached the accuracy of the full training with about 500 samples, a fifth of the original dataset size. Moreover, these methods led to better accuracies than with the training of the whole 2500 learning samples. The uncertainty methods RSAL and EQB resulted in poorer results.

Regarding LAI retrieval with GPR, similarly to the LCC case, the use of AL methods led to retrieval accuracies superior to that resulting from the full training dataset. The results converged faster here, with 400 samples needed to reach the full training accuracy. Random sampling led to similar results to the AL methods until 400 samples, but it then became unstable. Contrary to the above LCC case, RSAL performed better, leading to $R^2$ values similar to the other AL methods. In fact, all AL techniques showed similar accuracy results. The fastest converging method was PAL until 400 samples, but past this, CBD and EBD led to slightly superior accuracy results, although the overall difference between them is negligible.

\begin{figure}[H]
 \centering
	\footnotesize
	\begin{tabular}{ccc}
 & Samples at best result ($R^2$)  & Processing time (s) \\ 
\raisebox{1cm}{\rotatebox[origin=c]{90}{LCC}} & \IG[width=6.8cm]{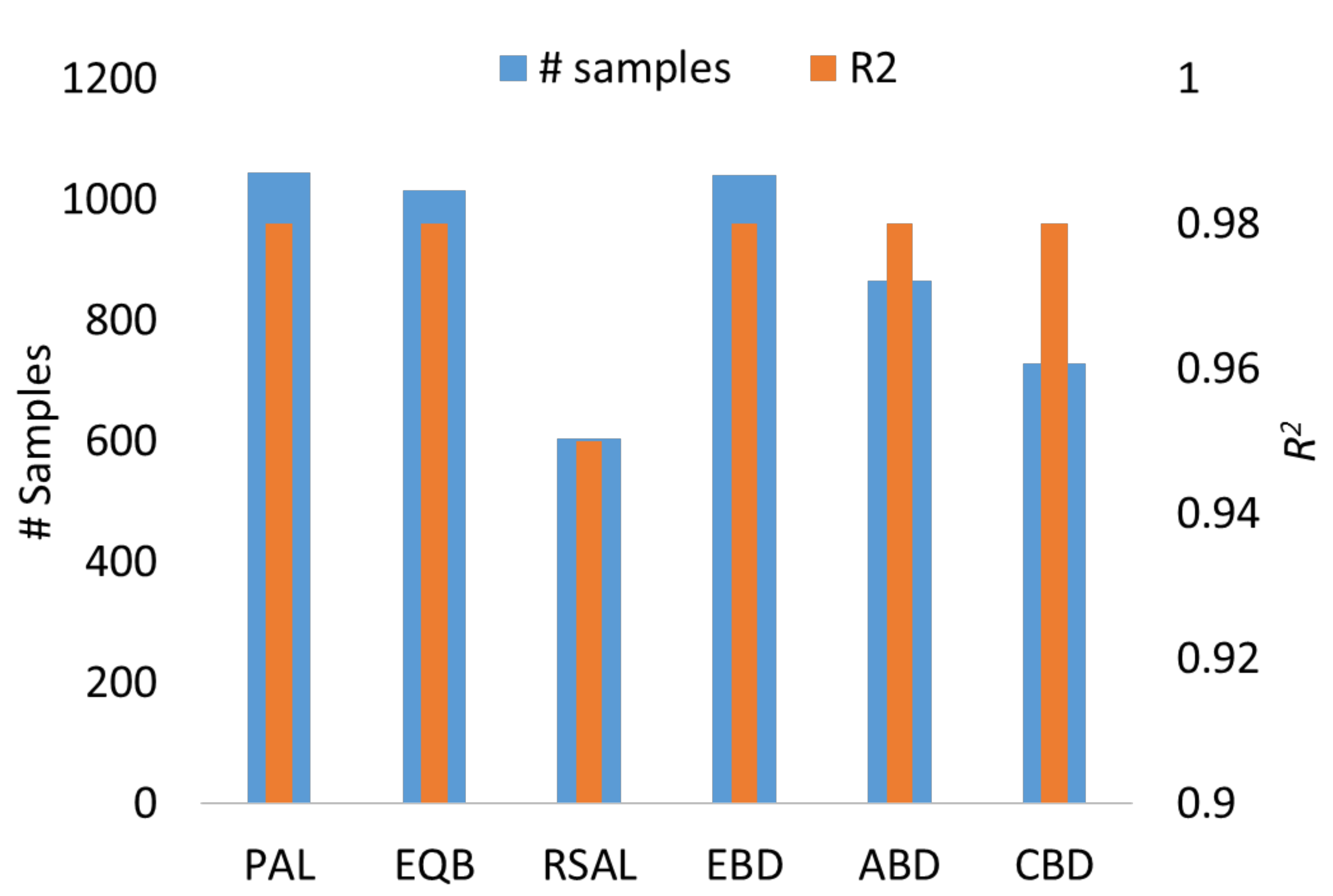} &
\IG[width=6.8cm]{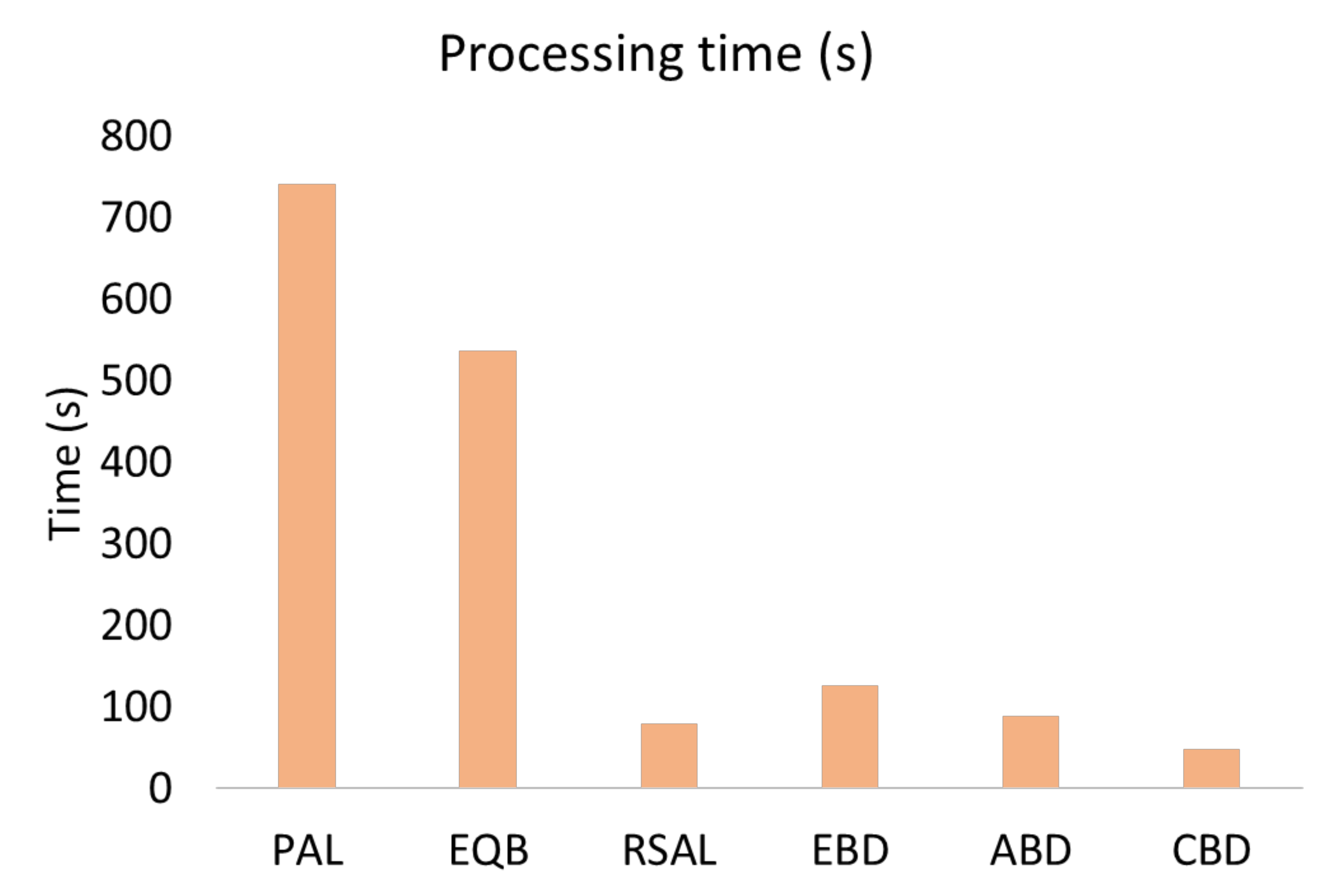} \\
\raisebox{1cm}{\rotatebox[origin=c]{90}{\hspace{0cm}LAI}} &
\IG[width=6.8cm]{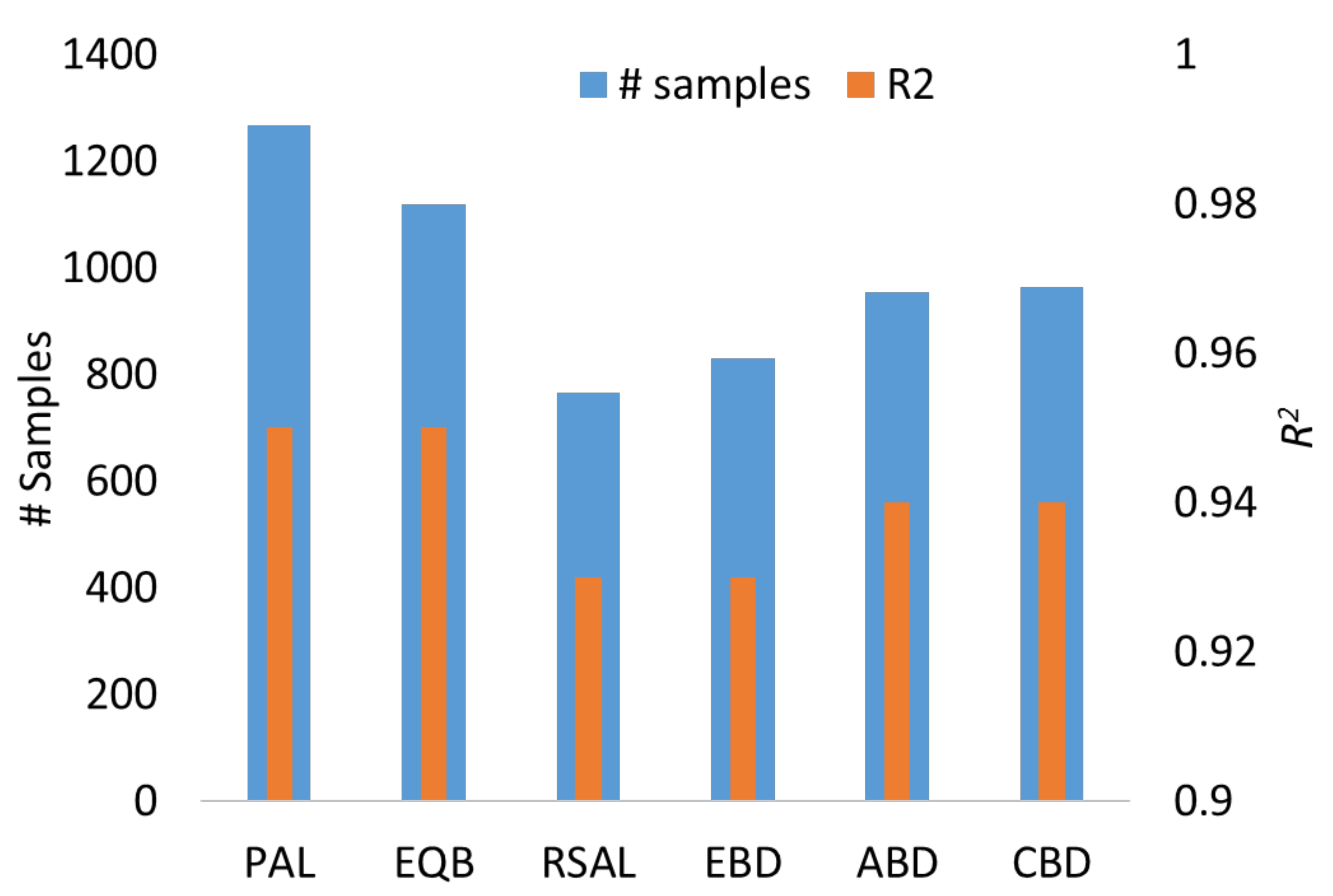} &
\IG[width=6.8cm]{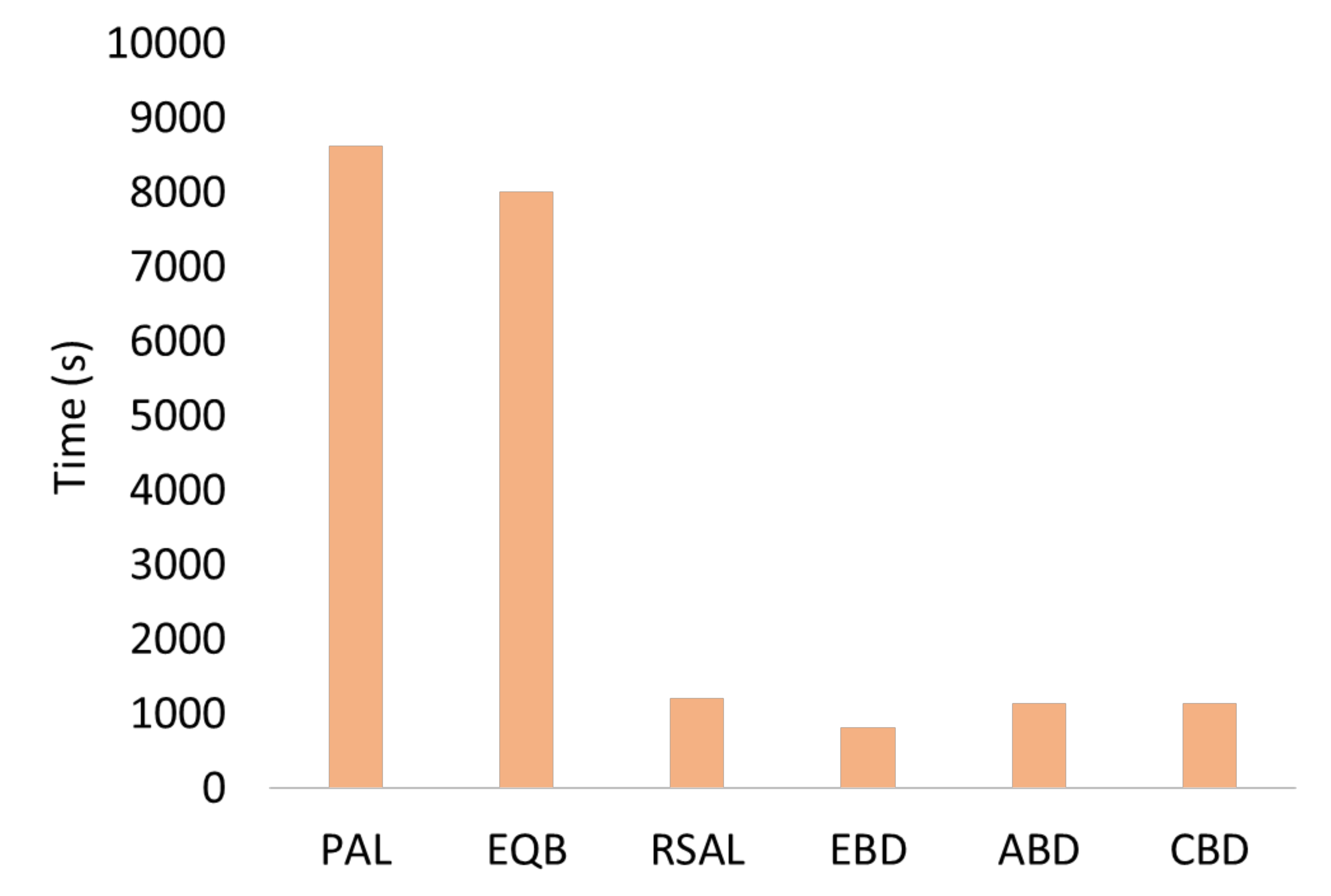} \\
\end{tabular}\vspace{12pt}
 \caption{The number of samples needed to reach maximal accuracies [left], and associated processing time [right] for LCC [top] and LAI [bottom].}
 \label{best_results}
\end{figure}

Finally, the number of samples needed to reach maximal retrieval accuracy and their corresponding processing time were recorded (Fig. \ref{best_results}). 
The diversity methods ABD and CBD processed fastest to reach maximized accuracies. Since these two diversity methods also led to high accuracies with relatively small training sets, they seem to be the best methods for operational applications. EBD and RSAL were also fast, but EBD performed unstable when comparing LCC against LAI, and RSAL led to poorest accuracies.  PAL and EQB, although yielding high accuracies, took the longest time to compute since these methods created five regression models to be compared.

\section{Conclusion and Outlook}\label{sec:conclude}

Kernel-based MLRAs allow for fast and accurate variable predictions, and they possess attractive properties to become the core of next-generation operational hybrid retrieval schemes~\cite{Verrelst2015b}. However most of them cannot process large training databases. For instance, here it was found that standard implementations of GPR cannot typically cope with more than 5000 samples, i.e. a biophysical variable associated with superspectral or hyperspectral data, under reasonable time, which hampers its direct application towards generic retrieval models. To mitigate this problem the use of six AL methods was explored in this letter. AL methods are promising to crystallize an optimal training sampling set that is manageable by kernel-based MLRA methods.

For two kernel-based MLRA methods, i.e. GPR and KRR, we demonstrated the utility of AL methods as an intelligent sampling step that selects the best possible samples from a large (simulated) training dataset. 
By using a data pool of PROSAIL-simulated LAI, LCC and S3-OLCI spectra, each AL method converged faster to the lower error bound than a random sampling strategy. Diversity criteria (EBD, ABD, CBD) performed generally fastest, both in terms of reaching high accuracies as well in processing time. Random sampling converged in the long run to similar accuracies but did not offer a stable error minimization over few samples. The reason each of the AL methods smoothly converged to the full training error bound is that the methods only allow the addition of samples that improve the overall accuracy of the model. These observations using simulated training data confirm earlier results that were obtained in experimental regression problem studies ~\cite{Pasolli2012,Douak2013}. This demands for a follow-up study to combine uncertainty and diversity criteria. Another attractive path to pursue is to select samples according to a density criterion \cite{Demir2014}.  
This work led to an updated ARTMO MLRA toolbox with the ability to apply AL methods into hybrid retrieval models in a semi-automatic manner. Combining both AL methods with machine learning regression algorithms makes the toolbox particularly powerful in processing optical remote sensing images into vegetation products. It opens possibilities towards 
the development of powerful (kernel-based) and generic hybrid retrieval schemes, e.g. based on RTM simulations. However, to pursue with developing hybrid strategies for an operational setting would imply introducing a larger training dataset, which requires more samples to be added per AL iteration. A generic validation dataset would then be required to ascertain the minimum amount of samples that stabilizes the retrieval performance. 
Another recommendation is to feed the AL methods not only with simulated data, but with data collected in field or coming from earlier products. For instance, ESA's current LAI product comes from a NN that is trained by earlier LAI products \cite{Baret2013}. Also, while the initial training set was chosen in a random way, a more sophisticated initialization strategy may further improve the AL sequence. Eventually, the freely available MLRA toolbox will allow users to develop their own robust, compact, accurate, and generically applicable retrieval algorithms with little user interaction.

\end{document}